\documentclass[showpacs,preprintnumbers,amsmath,amssymb,twocolumn]{revtex4}

\usepackage{graphicx}

\newcommand{\lsm}{La$_{0.7}$Sr$_{0.3}$MnO$_3 $}
\newcommand{\feo}{Fe$_3$O$_4 $}

\begin{document}


\title{Giant Magnetoresistance in an all-oxide spacerless junction}

\author{M. P. Singh}
\author{B. Carvello}

\author{L. Ranno}
 \email{laurent.ranno@grenoble.cnrs.fr}
\affiliation{Laboratoire Louis N\'eel, CNRS-UJF-INPG, Polygone CNRS, BP 166, 38042 Grenoble cedex 09, France \\
}

\date{10 July 2006}

\begin{abstract}
  We report the fabrication of an oxide-specific type of magnetoresistive
  junction, which is a ferromagnetic bilayer. Both
  electrodes are high spin-polarization oxides: magnetite (\feo) and manganite
  (\lsm). Negligible magnetic coupling between both ferromagnetic electrodes is
  realised, which allows to obtain parallel and antiparallel magnetic
  configurations of the electrodes when sweeping the applied magnetic field. The
  structure exhibits negative giant magnetoresistance (GMR) at low temperatures.
  This negative MR shows that both electrodes stay spin-polarized at the interface
  and have opposite spin polarizations, \emph{i.e.} the \feo\ layer has a negative
  spin polarization at low temperature. Maximum GMR (-5~\%) is obtained at 55~K.
\end{abstract}


\maketitle

\section{Introduction}

Highly spin-polarized ferromagnetic oxides, such as CrO$_2$, \feo, \lsm\ (LSMO)
have been the focus of recent fundamental and technological studies in the field
of spin electronics. Using these materials, various devices, such as giant
magnetoresistance (GMR) junctions \cite{vandijken04} and tunnel
magnetoresistance (TMR) junctions \cite{seneor99} have been fabricated and
studied. To fabricate a magnetoresistive device based on a junction, usually two
ferromagnetic layers are separated by a thin non magnetic layer (the spacer).
The nature of the spacer is chosen in order to control the spin-dependent
transport mechanism at the interface: metallic spacer (GMR) or insulating spacer
(TMR).

In the usual case of transition metal electrodes, the thickness of the spacer is
chosen in order to magnetically decouple the magnetic layers (\emph{i.e.}
thickness larger than a few atomic planes to break the direct coupling exchange
path, and to prevent indirect coupling such as the RKKY one). The transport
across the spacer must conserve the spin information, thus
the spacer thickness must be kept thinner
than a few mean free paths (current-in-plane cip-GMR) or spin diffusion lengths
(current-perpendicular-to-plane cpp-GMR) or a few 1/$\rm{k_F}$ (tunnel
probability in TMR junction).

A spacer is not necessary if it is possible to weaken the magnetic coupling
between both electrodes. Such devices have already been proposed: ballistic
junctions \cite{garcia99} or break junctions \cite{gabureac04}. These junctions
were designed to break the exchange coupling between two transition metals. To
achieve this, both electrodes have to be mechanically separated, which is a
difficult step, source of non reproducibilities, and sensitive to parasitic
phenomena such as magnetostriction \cite{egelhoff04}. In this paper we propose a
simple solid state structure, adapted to collective fabrication.

In oxides, magnetic coupling is due to indirect exchange (3d ion - oxygen - 3d
ion) and it is very sensitive to the atomic details of such a bond. For example
it is possible to weaken the coupling by changing the bond angle (manganite's
$\rm{T_C}$ varies as a function of the Mn-O-Mn bond angle \cite{hwang95}). Thus,
tuning the interfacial magnetic coupling is achievable in oxides.

The spin-polarization of a material is positive if the majority spin at the
Fermi level is parallel to the magnetization and negative if the minority spin
at the Fermi level is parallel to the magnetization. Half metallic ferromagnets
have a spin polarization of 100~\% (only one spin direction is present at the
Fermi level). Magnetite (\feo) stands out as a predicted half metallic
ferromagnet (ferrimagnet in fact) with negative spin polarization
\cite{yanase84} and a remarkably high Curie temperature ($\rm{T_C}$) of 858~K.
\lsm\ is predicted to have 100~\% positive polarization \cite{livesay99} with
$\rm{T_C}$ of 350~K. A junction between two such half metallic ferromagnetic
compounds would in theory behave as an ideal magnetic-field-controlled switch with 100~\%
negative magnetoresistance (MR).

As mentioned above, recent efforts have been made to fabricate \feo/I/\lsm\
junctions, where I is SrTiO$_3$, CoCr$_2$O$_4$ \cite{ogale98,hu03}, 
but a \feo/\lsm\ junction without a spacer has never been proposed.

\section{Experimental Details}

The \feo/\lsm\ bilayers were grown on (001)-oriented SrTiO$_3$ (STO) substrates using pulsed
laser deposition. First, the LSMO layer was grown at 1023 K under 40 Pa of
O$_2$, then the \feo\ layer was grown at 623~K under 5.10$^{-4}$ Pa of O$_2$.
The thickness of LSMO was 50 nm whereas the \feo\ was grown with two different
thicknesses, 15 nm and 50 nm, estimated \textit{in situ} by optical
reflectometry. Prior to the deposition, the substrate was heated in oxygen up to
the deposition temperature.

To study the magnetotransport properties of the junction, 50~nm Au was deposited
upon the 15/50~nm \feo/LSMO/STO structure at room temperature using the
sputtering technique and subsequently, junctions of 500 $\times$~500~$\mu$m and 140 
$\times$~140~$\mu$m were fabricated by photolithography and Ar ion etching process. All
transport measurements were carried out with cpp geometry and applied magnetic
field parallel to the plane.

X-ray diffraction (XRD) study was carried out to examine the structural
properties of the bilayers. Despite the 1~\% lattice mismatch between LSMO and
STO, the LSMO growth is pseudomorphic up to a critical thickness (100 nm) larger
than the thickness chosen for these bilayers \cite{ranno02}. The LSMO film is
epitaxially-strained (0.18$^{\circ}$ FWHM rocking-curve) and a large
epitaxially-induced magneto-elastic anisotropy is present \cite{ranno02}. The
\feo\ film on LSMO is textured with multiple orientations (diffraction peaks
corresponding to [001] and [011] directions, but not [311]), whereas \feo\ films
on SrTiO$_3$ deposited under similar conditions, were grown textured along the
[001] direction with 1$^{\circ}$ FWHM rocking curve. The details of the
deposition of films of LSMO and \feo\ on STO have been reported
elsewhere~\cite{ranno02, carvello04}.

\section{Results}

To check for magnetic coupling between both oxide layers, M(H) hysteresis loops
of the unpatterned 50/50~nm bilayer structure were measured in the temperature
range 10-350~K and up to 3 Tesla using a VSM and a SQUID magnetometer.

\begin{figure}[b!]
\includegraphics[width=9cm]{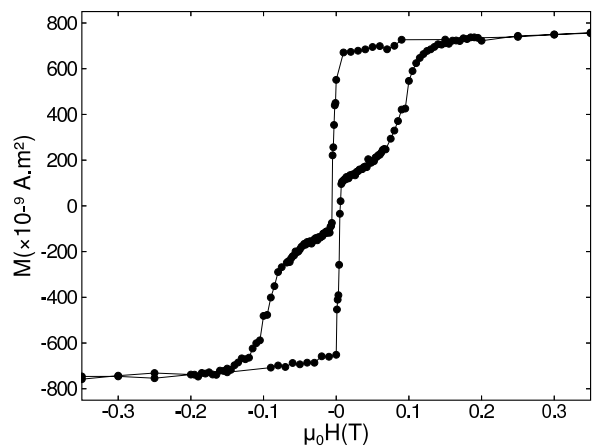}
\caption{\label{fig:MH}Hysteresis loops of unpatterned bilayer film at 50~K}
\end{figure}

Fig.~1
shows the typical hysteresis loop from an unpatterned bilayer
structure at 50~K, with the magnetic field applied along the substrate [110]
axis, which is the easy direction of the \lsm\ layer. The hysteresis loop clearly
shows two distinct coercive fields (about 5~mT and 100~mT), which correspond to
LSMO and \feo, respectively. The temperature dependence of \feo\ coercivity
shows the Verwey transition at 110~K ($\rm{T_V}$~=~122~K in single crystals).

To study the magnetic coupling between these layers, minor loops of the softer
layer (LSMO) were measured at 10~K. No shift of the loops was detected.
Thus no exchange bias field larger than 3~mT exists.
The large coercivity difference between layers and squareness of
the LSMO hysteresis loop create well-defined parallel and antiparallel magnetic
configurations.

As far as transport is concerned, we have measured I(V) characteristics from 5
to 300~K. They are linear up to the point where heating effects come into play.
The evolution of the resistance with temperature
(Fig.~2) 
can be
divided into 3 regimes. The high temperature regime, above 90~K, exhibits the
well-known resistance and CMR (colossal magnetoresistance) of manganites, and
can thus be attributed to the LSMO electrode, which dominates the transport at
these temperatures due to geometrical reasons. 
Between 30 and 90~K, the
transport is thermally activated, due to the increasing \feo\ dominance when
temperature decreases. Below 30~K, a plateau is observed, which is surprising in
a \feo-dominated regime. This remains to be investigated, but could be explained
by the onset of a hot electron transport mechanism due to the high electric
field (40~kV/cm), such as the one observed in \cite{chern92}.

\begin{figure}[t!]
\includegraphics[width=9cm]{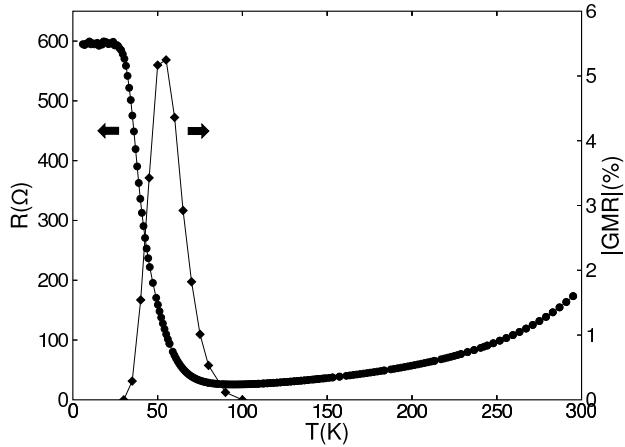}
\caption{\label{fig:RT} Resistance and GMR of a 140 $\times$~140~$\mu$m junction
measured with a current of 100~$\mu$A}
\end{figure}

At high field (1 to 6~T range), the junction shows a negative
magnetoresistance. This high field MR is large below 40~K (over -1~\%/T,
consistent with \feo\ thin films), smaller between 40 and 100~K, and increases 
to high values (-2~\%/T) at high temperature (LSMO CMR). In the intermediate
regime, though, this negative slope is only visible above 2~T. Under that field,
\feo\ is poorly saturated, and GMR dominates (see below), giving a positive
slope.

\begin{figure}[b!]
\includegraphics[width=9cm]{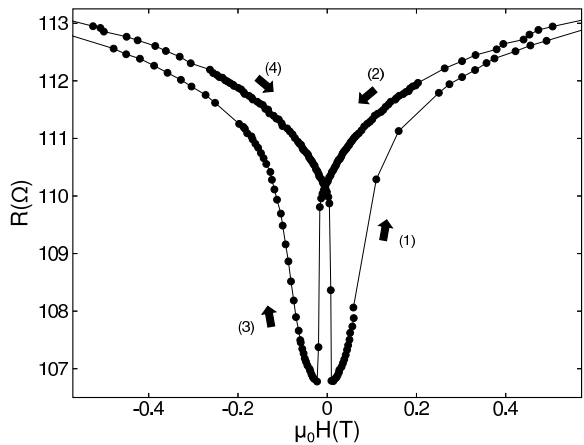}
\caption{\label{fig:GMR} R(H) magnetotransport measurement of a 140 $\times$ 140
$\mu$m junction for T~=~55~K }
\end{figure}

In the intermediate temperature range, the magnetic field dependent transport
measurement shows a characteristic inverse GMR behavior 
(Fig.~3, the applied field is sweeped as the arrows indicate).
The magnetic fields at which the junction resistance changes the most abruptly
correspond to the coercivities of both oxide layers and the junction resistance
is lower when the magnetizations of both layers are antiparallel to each other.
The GMR was measured at $\pm$~100~$\mu$A constant current as a function of
temperature 
(Fig.~2). 
$R_{\uparrow \downarrow}$ is measured at 20~mT
and $R_{\uparrow \uparrow}$ at 400~mT. This magnetoresistance diminishes in
absolute value both above 90~K and below 30~K. The maximum GMR of -~5.2~\% is
found at 55~K.

\section{Discussion}

The two keypoints which have to be discussed are the origin of the magnetic
decoupling of the \lsm\ and \feo\ electrodes and the transport mechanism
responsible for the large magnetoresistance. The ferromagnetic coupling
mechanism is double exchange (Mn$^{3+}$-O-Mn$^{4+}$ bonds) in the case of LSMO,
and is superexchange (Fe$^{3+}$(A)-O-Fe$^{2+/3+}$(B) bonds) as well as double
exchange (Fe$^{3+}$(B)-O-Fe$^{2+}$(B) bonds) in the case of \feo. To
magnetically decouple the two layers, these nearest-neighbour mechanisms have to
be weakened. The interface between LSMO and \feo\ is a structurally
disordered layer due to the 6.7~\% lattice mismatch, which prevents
heteroepitaxy. In oxides, due to the localised character of electrons, weakening
the exchange is much easier than in transition metals where electrons are more
delocalised (RKKY coupling range can reach a few nanometers). So one disordered
layer due to the lattice mismatch
between a perovskite and a spinel ferromagnet is enough to reduce the exchange
coupling and to decouple both layers. This is a general statement since the
lattice mismatch between a spinel and a perovskite structure will always be a
few~\%.

We claim that our structure is a bilayer. However the presence of an intermixed
layer between the electrodes has to be ruled out to support this claim. In our
system any intermixing layer would be made of Fe and Mn ions and therefore it
would be magnetic. Since transport in conducting ferromagnetic oxides is based
on a nearest-neighbour hopping mechanism, any magnetic layer depolarizes the
current. Thus, the characteristic type of magnetoresistance we can measure rules out the presence
of an intermixing layer. Furthermore since the \feo\ layer is deposited at low
temperature (623~K), the spinel/perovskite interface is expected to be stable.
We have recently conducted a TEM study on the SrTiO$_3$~/~\feo\ interface, which
showed that the perovskite / spinel mismatch can be accomodated through a
regular array of dislocations \cite{carvello05} and confirms that no intermixing takes place.

As the MR is small (-5~\%), spin disorder in the interfacial plane, leading to a
partial depolarisation, cannot be ruled out. However, this cannot be called a
distinct magnetic layer, it is better characterized as an interfacial disorder.

The two electrodes are in direct electrical contact, so the nature of the
transport mechanism is related to the presence or absence of an electrical
barrier. The cpp transport characteristics are ohmic, and the R(T) exhibits no
regime that could be interpreted as a tunnel transport.

Since a significant MR has been measured and the characteristic fields of this MR are
the coercive fields of both electrodes a spin-coherent mechanism has to be
proposed. Through the interface, transport could still be based on a hopping
mechanism. Interface disorder only reduces hopping integrals (also impacting magnetism by
reducing the double-exchange coupling). Thus the MR mechanism is closer to the cpp-GMR
than to the TMR mechanism and the only significant interfacial resistance is the
GMR itself.

The value of the magnetoresistance is difficult to interpret in a quantitative
manner, given that the resistance of the interface is not dominant compared to
that of the electrodes. Thus the -5.2~\% GMR ratio is not intrinsic, and could
be enhanced through an optimized junction pattern.

It is also worth noting
that it is difficult to obtain a real parallel or antiparallel state with \feo.
Because of structural defects present in all \feo\ thin films (antiphase
boundaries), the remanence is less than 80~\%, and full saturation is not
achieved even at large fields. Lastly, the decay of the GMR at high temperature
can be explained by the decrease of LSMO polarisation well below $\rm{T_C}$, as
is known from other studies (\cite{favre01}). Nonetheless, our measurements
evidence negative spin polarization of \feo, as predicted, down to the interface
with LSMO.

For these reasons, we propose that the mechanism responsible for the exchange
weakening and the large MR in our bilayer is fundamentally different from the
TMR and spacer-assisted cpp-GMR mecanisms reported in other studies. It is
purely interfacial (i.e. spacerless) and specific to oxides.


\begin{acknowledgments}
One of the authors (MPS) would like to thank EGIDE and the French Foreign
Ministry, for providing the postdoctoral fellowship.
\end{acknowledgments}


%
%
%
%
%
%
%


\begin{thebibliography}{15}
\expandafter\ifx\csname natexlab\endcsname\relax\def\natexlab#1{#1}\fi
\expandafter\ifx\csname bibnamefont\endcsname\relax
  \def\bibnamefont#1{#1}\fi
\expandafter\ifx\csname bibfnamefont\endcsname\relax
  \def\bibfnamefont#1{#1}\fi
\expandafter\ifx\csname citenamefont\endcsname\relax
  \def\citenamefont#1{#1}\fi
\expandafter\ifx\csname url\endcsname\relax
  \def\url#1{\texttt{#1}}\fi
\expandafter\ifx\csname urlprefix\endcsname\relax\def\urlprefix{URL }\fi
\providecommand{\bibinfo}[2]{#2}
\providecommand{\eprint}[2][]{\url{#2}}

\bibitem[{\citenamefont{van Dijken et~al.}(2004)\citenamefont{van Dijken, Fain,
  Watts, and Coey}}]{vandijken04}
\bibinfo{author}{\bibfnamefont{S.}~\bibnamefont{van Dijken}},
  \bibinfo{author}{\bibfnamefont{X.}~\bibnamefont{Fain}},
  \bibinfo{author}{\bibfnamefont{S.~M.} \bibnamefont{Watts}}, \bibnamefont{and}
  \bibinfo{author}{\bibfnamefont{J.~M.~D.} \bibnamefont{Coey}},
  \bibinfo{journal}{Phys. Rev. B 70} \textbf{\bibinfo{volume}{70}},
  \bibinfo{pages}{052409} (\bibinfo{year}{2004}).

\bibitem[{\citenamefont{Seneor et~al.}(1999)\citenamefont{Seneor, Fert,
  Maurice, Montaigne, Petroff, and Vaur\`es}}]{seneor99}
\bibinfo{author}{\bibfnamefont{P.}~\bibnamefont{Seneor}},
  \bibinfo{author}{\bibfnamefont{A.}~\bibnamefont{Fert}},
  \bibinfo{author}{\bibfnamefont{J.-L.} \bibnamefont{Maurice}},
  \bibinfo{author}{\bibfnamefont{F.}~\bibnamefont{Montaigne}},
  \bibinfo{author}{\bibfnamefont{F.}~\bibnamefont{Petroff}}, \bibnamefont{and}
  \bibinfo{author}{\bibfnamefont{A.}~\bibnamefont{Vaur\`es}},
  \bibinfo{journal}{Appl. Phys. Lett.} \textbf{\bibinfo{volume}{74}},
  \bibinfo{pages}{4017} (\bibinfo{year}{1999}).

\bibitem[{\citenamefont{Garcia et~al.}(1999)\citenamefont{Garcia, Mu\~noz, and
  Zhao}}]{garcia99}
\bibinfo{author}{\bibfnamefont{N.}~\bibnamefont{Garcia}},
  \bibinfo{author}{\bibfnamefont{M.}~\bibnamefont{Mu\~noz}}, \bibnamefont{and}
  \bibinfo{author}{\bibfnamefont{Y.-W.} \bibnamefont{Zhao}},
  \bibinfo{journal}{Phys. Rev. Lett.} \textbf{\bibinfo{volume}{82}},
  \bibinfo{pages}{2923} (\bibinfo{year}{1999}).

\bibitem[{\citenamefont{Gabureac et~al.}(2004)\citenamefont{Gabureac, Viret,
  Ott, and Fermon}}]{gabureac04}
\bibinfo{author}{\bibfnamefont{M.}~\bibnamefont{Gabureac}},
  \bibinfo{author}{\bibfnamefont{M.}~\bibnamefont{Viret}},
  \bibinfo{author}{\bibfnamefont{F.}~\bibnamefont{Ott}}, \bibnamefont{and}
  \bibinfo{author}{\bibfnamefont{C.}~\bibnamefont{Fermon}},
  \bibinfo{journal}{Phys. Rev. B} \textbf{\bibinfo{volume}{69}},
  \bibinfo{pages}{100401(R)} (\bibinfo{year}{2004}).

\bibitem[{\citenamefont{Egelhoff et~al.}(2004)\citenamefont{Egelhoff, Gan,
  Ettedgui, Kadmon, Powell, Chen, Shapiro, McMichael, Mallett, Moffat,
  Stiles, and Svedberg}}]{egelhoff04}
\bibinfo{author}{\bibfnamefont{W.~F.} \bibnamefont{Egelhoff}},
  \bibinfo{author}{\bibfnamefont{L.}~\bibnamefont{Gan}},
  \bibinfo{author}{\bibfnamefont{H.}~\bibnamefont{Ettedgui}},
  \bibinfo{author}{\bibfnamefont{Y.}~\bibnamefont{Kadmon}},
  \bibinfo{author}{\bibfnamefont{C.~J.} \bibnamefont{Powell}},
  \bibinfo{author}{\bibfnamefont{P.~J.} \bibnamefont{Chen}},
  \bibinfo{author}{\bibfnamefont{A.~J.} \bibnamefont{Shapiro}},
  \bibinfo{author}{\bibfnamefont{R.~D.} \bibnamefont{McMichael}},
  \bibinfo{author}{\bibfnamefont{J.~J.} \bibnamefont{Mallett}},
  \bibinfo{author}{\bibfnamefont{T.~P.} \bibnamefont{Moffat}},
  \bibinfo{author}{\bibfnamefont{M.~D.} \bibnamefont{Stiles}}, \bibnamefont{and}
  \bibinfo{author}{\bibfnamefont{E.~B.} \bibnamefont{Svedberg}},
  \bibinfo{journal}{J. Appl. Phys.}
  \textbf{\bibinfo{volume}{95}}, \bibinfo{pages}{7554} (\bibinfo{year}{2004}).

\bibitem[{\citenamefont{Hwang et~al.}(1995)\citenamefont{Hwang, Cheong,
  Radaelli, Marezio, and Batlogg}}]{hwang95}
\bibinfo{author}{\bibfnamefont{H.~Y.} \bibnamefont{Hwang}},
  \bibinfo{author}{\bibfnamefont{S.-W.} \bibnamefont{Cheong}},
  \bibinfo{author}{\bibfnamefont{P.~G.} \bibnamefont{Radaelli}},
  \bibinfo{author}{\bibfnamefont{M.}~\bibnamefont{Marezio}}, \bibnamefont{and}
  \bibinfo{author}{\bibfnamefont{B.}~\bibnamefont{Batlogg}},
  \bibinfo{journal}{Phys. Rev. Lett.} \textbf{\bibinfo{volume}{75}},
  \bibinfo{pages}{914} (\bibinfo{year}{1995}).

\bibitem[{\citenamefont{Yanase and Siratori}(1984)}]{yanase84}
\bibinfo{author}{\bibfnamefont{A.}~\bibnamefont{Yanase}} \bibnamefont{and}
  \bibinfo{author}{\bibfnamefont{K.}~\bibnamefont{Siratori}},
  \bibinfo{journal}{J. Phys. Soc. Jap.} \textbf{\bibinfo{volume}{53}},
  \bibinfo{pages}{312} (\bibinfo{year}{1984}).

\bibitem[{\citenamefont{Livesay et~al.}(1999)\citenamefont{Livesay, West,
  Dugdale, Santi, and Jarlborg}}]{livesay99}
\bibinfo{author}{\bibfnamefont{E.~A.} \bibnamefont{Livesay}},
  \bibinfo{author}{\bibfnamefont{R.~N.} \bibnamefont{West}},
  \bibinfo{author}{\bibfnamefont{S.~B.} \bibnamefont{Dugdale}},
  \bibinfo{author}{\bibfnamefont{G.}~\bibnamefont{Santi}}, \bibnamefont{and}
  \bibinfo{author}{\bibfnamefont{T.}~\bibnamefont{Jarlborg}},
  \bibinfo{journal}{J. Phys.: Condens Matter} \textbf{\bibinfo{volume}{11}},
  \bibinfo{pages}{L279} (\bibinfo{year}{1999}).

\bibitem[{\citenamefont{Ogale et~al.}(1998)\citenamefont{Ogale, Ghosh, Pai,
  Robson, Li, Jin, Greene, Ramesh, Venketesan, and Johnson}}]{ogale98}
\bibinfo{author}{\bibfnamefont{S.~B.} \bibnamefont{Ogale}},
  \bibinfo{author}{\bibfnamefont{K.}~\bibnamefont{Ghosh}},
  \bibinfo{author}{\bibfnamefont{S.~P.} \bibnamefont{Pai}},
  \bibinfo{author}{\bibfnamefont{M.}~\bibnamefont{Robson}},
  \bibinfo{author}{\bibfnamefont{E.}~\bibnamefont{Li}},
  \bibinfo{author}{\bibfnamefont{I.}~\bibnamefont{Jin}},
  \bibinfo{author}{\bibfnamefont{R.~L.} \bibnamefont{Greene}},
  \bibinfo{author}{\bibfnamefont{R.}~\bibnamefont{Ramesh}},
  \bibinfo{author}{\bibfnamefont{T.}~\bibnamefont{Venketesan}},
  \bibnamefont{and} \bibinfo{author}{\bibfnamefont{M.}~\bibnamefont{Johnson}},
  \bibinfo{journal}{Mater. Sci. Eng. B} \textbf{\bibinfo{volume}{56}},
  \bibinfo{pages}{134 } (\bibinfo{year}{1998}).

\bibitem[{\citenamefont{Hu et~al.}(2003)\citenamefont{Hu, Chopdekar, and
  Suzuki}}]{hu03}
\bibinfo{author}{\bibfnamefont{G.}~\bibnamefont{Hu}},
  \bibinfo{author}{\bibfnamefont{R.}~\bibnamefont{Chopdekar}},
  \bibnamefont{and} \bibinfo{author}{\bibfnamefont{Y.}~\bibnamefont{Suzuki}},
  \bibinfo{journal}{J. Appl. Phys.} \textbf{\bibinfo{volume}{93}},
  \bibinfo{pages}{7516} (\bibinfo{year}{2003}).

\bibitem[{\citenamefont{Ranno et~al.}(2002)\citenamefont{Ranno, Llobet, Tiron,
  and Favre-Nicolin}}]{ranno02}
\bibinfo{author}{\bibfnamefont{L.}~\bibnamefont{Ranno}},
  \bibinfo{author}{\bibfnamefont{A.}~\bibnamefont{Llobet}},
  \bibinfo{author}{\bibfnamefont{R.}~\bibnamefont{Tiron}}, \bibnamefont{and}
  \bibinfo{author}{\bibfnamefont{E.}~\bibnamefont{Favre-Nicolin}},
  \bibinfo{journal}{Appl. Surf. Sc.} \textbf{\bibinfo{volume}{188}},
  \bibinfo{pages}{170} (\bibinfo{year}{2002}).

\bibitem[{\citenamefont{Carvello and Ranno}(2004)}]{carvello04}
\bibinfo{author}{\bibfnamefont{B.}~\bibnamefont{Carvello}} \bibnamefont{and}
  \bibinfo{author}{\bibfnamefont{L.}~\bibnamefont{Ranno}}, \bibinfo{journal}{J.
  Magn. Magn. Mater.} \textbf{\bibinfo{volume}{272-276}}, \bibinfo{pages}{1926}
  (\bibinfo{year}{2004}).

\bibitem[{\citenamefont{Chern et~al.}(1992)\citenamefont{Chern, Berry, Lind,
  Mathias, and Testardi}}]{chern92}
\bibinfo{author}{\bibfnamefont{G.}~\bibnamefont{Chern}},
  \bibinfo{author}{\bibfnamefont{S.~D.} \bibnamefont{Berry}},
  \bibinfo{author}{\bibfnamefont{D.~M.} \bibnamefont{Lind}},
  \bibinfo{author}{\bibfnamefont{H.}~\bibnamefont{Mathias}}, \bibnamefont{and}
  \bibinfo{author}{\bibfnamefont{L.~R.} \bibnamefont{Testardi}},
  \bibinfo{journal}{Phys. Rev. B} \textbf{\bibinfo{volume}{45}},
  \bibinfo{pages}{3644 } (\bibinfo{year}{1992}).

\bibitem[{\citenamefont{Carvello et~al.}()\citenamefont{Carvello, Warin, and
  Ranno}}]{carvello05}
\bibinfo{author}{\bibfnamefont{B.}~\bibnamefont{Carvello}},
  \bibinfo{author}{\bibfnamefont{P.}~\bibnamefont{Warin}}, \bibnamefont{and}
  \bibinfo{author}{\bibfnamefont{L.}~\bibnamefont{Ranno}}, \bibinfo{note}{to be
  published}.

\bibitem[{\citenamefont{Favre-Nicolin et~al.}(2001)\citenamefont{Favre-Nicolin,
  Ranno, Dubourdieu, and Rosina}}]{favre01}
\bibinfo{author}{\bibfnamefont{E.}~\bibnamefont{Favre-Nicolin}},
  \bibinfo{author}{\bibfnamefont{L.}~\bibnamefont{Ranno}},
  \bibinfo{author}{\bibfnamefont{C.}~\bibnamefont{Dubourdieu}},
  \bibnamefont{and} \bibinfo{author}{\bibfnamefont{M.}~\bibnamefont{Rosina}},
  \bibinfo{journal}{Thin Solid Films} \textbf{\bibinfo{volume}{400}},
  \bibinfo{pages}{165 } (\bibinfo{year}{2001}).

\end{thebibliography}
\end{document}